# Collective Diffusion Over Networks: Models and Inference

**Akshat Kumar**
Analytics and Optimization
IBM Research India
akkumaro@in.ibm.com

**Daniel Sheldon**
Dept. of Computer Science
University of Massachusetts Amherst
sheldon@cs.umass.edu

**Biplav Srivastava**
Analytics and Optimization
IBM Research India
sbiplav@in.ibm.com

## Abstract

Diffusion processes in networks are increasingly used to model the spread of information and social influence. In several applications in computational sustainability such as the spread of wildlife, infectious diseases and traffic mobility pattern, the observed data often consists of only *aggregate* information. In this work, we present new models that generalize standard diffusion processes to such collective settings. We also present optimization based techniques that can accurately learn the underlying dynamics of the given contagion process, including the hidden network structure, by only observing the time a node becomes active and the associated aggregate information. Empirically, our technique is highly robust and accurately learns network structure with more than 90% recall and precision. Results on real-world flu spread data in the US confirm that our technique can also accurately model infectious disease spread.

## 1 Introduction

Dynamic phenomena such as the spread of information, ideas, and opinions (Domingos and Richardson, 2001; Kempe *et al.*, 2003; Leskovec *et al.*, 2007) can be described as a *diffusion* process or *cascade* over an underlying network. Similar diffusion processes have also been used for *metapopulation* modeling in the ecology literature to describe how wildlife spreads over a fragmented landscape (Hanski, 1999) and to describe infectious disease propagation among humans (Anderson and May, 2002; Halloran *et al.*, 2010). Such models are crucial for several decision making problems in computational sustainability such as in spatial conservation planning that addresses the question of how to allocate resources to maximize the population spread of an endangered species over a period of time (Sheldon *et al.*, 2010; Ahmadizadeh *et al.*, 2010; Golovin *et al.*, 2011; Kumar *et al.*, 2012).

A fundamental problem in using such diffusion-based models is the estimation of parameters, including the hidden network structure, that govern the contagion process. Recent progress had been made in learning the diffusion parameters of social networks (Myers and Leskovec, 2010; Gomez-Rodriguez *et al.*, 2012; Netrapalli and Sanghavi, 2012; Wang *et al.*, 2012). Myers and Leskovec (2010) formulate the problem of network structure learning as a separable convex program. Gomez-Rodriguez *et al.* (2012) address the problem using submodular optimization. Netrapalli and Sanghavi (2012) address the complementary question of how many observed cascades are necessary to correctly learn the structure of a network. Wang *et al.* (2012) enrich the structure learning problem using additional features from Twitter data.

An implicit assumption commonly made in previous approaches in the social network setting is that one can track each individual in the network and exploit this information during inference. However, in several computational sustainability domains such as ecology, social sciences and transportation, data identifying a single individual is rarely available. For example, for population modeling of migratory birds, one may only know the total number of birds present in a geographical area. While modeling the spread of infectious diseases, one may only know the total number of infected individuals in a community. Similarly, traffic data usually takes the form of vehicle counts leaving or entering an intersection. In theory, one can model such aggregate behavior by explicitly reasoning about each individual in the population. However, such a model cannot be scaled to large population sizes.

We therefore present new *collective diffusion* models that can reason with the aggregate data without the need to model individual-level behavior. These models generalize the well known diffusion models such as the independent cascade model (Kempe *et al.*, 2003). We show how such models can be used to analyze the population dynamics of wildlife and spread of contagious diseases. We also present a model that can address collective diffusion in domains such as transportation that do not fall under the independent cascade model. We highlight how this model is closely related to the recently developed class of collective graphical models (CGMs) (Sheldon and Dietterich, 2011). Such connections are attractive as they open the door to the application of efficient inference techniques developed for CGMs to transportation domain.

The primary algorithmic contribution of our work is to develop algorithms for learning in collective diffusion models using observed data about the infection times of nodes and the associated aggregate information. Using scalable techniques based on convex optimization, we show that our approach can accurately learn network structure with more than 90% precision and recall on large synthetic benchmarks even with a limited number of observed cascades. Furthermore, our approach can also learn edge strength parameters accurately, with less than 2% error. Results on real-world flu spread data available from Centers for Disease Control and Prevention (CDC) in the US confirm that our technique can also accurately model infectious disease spread.

## 2 Diffusion Over Networks

We first introduce the well known *independent cascade* model, also called the *Susceptible-Infected* (SI) model, for diffusion over a network (Kempe *et al.*, 2003) and later present its collective variants. The main steps in this model are the following:

- We start with an initial set $S_0$ of active nodes called *seeds* in the network. The process then unfolds in discrete time steps.
- When a node $v$ first becomes active, it is given a single chance to activate each of its currently inactive neighbors $w$. It succeeds with probability $p_{vw}$ independently of the history of activations so far. Whether the node $v$ succeeds or fails in activating the node $w$, it cannot make further attempts to activate $w$ in the future.
- If a currently inactive node $w$ has multiple newly active neighbors, their activation attempts of $w$ are sequenced arbitrarily.
- A node $w$, once active, remains active for the entire diffusion process.

There are several extensions of the basic diffusion model above such as those allowing the nodes to re-cover and become infected again. It is easy to fold such extensions into the basic SI model using a time-indexed layered graph (Kempe *et al.*, 2003). A crucial inference problem in such a setting is estimating the edge activation probabilities $p_{vw}$. The edge activation probabilities also identify the connectivity structure of the network—if $p_{vw}=0$, then there is no directed edge from node $v$ to $w$. Next, we describe the observation model commonly used to address this parameter learning problem.

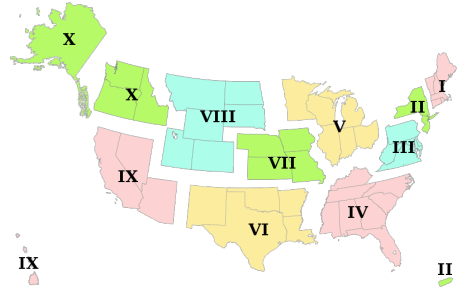

Figure 1: The US map showing 10 federal regions

**Observation model:** A cascade $c$ over such a network starts with a set of initially active nodes at time $t=0$. As the cascade progresses in discrete time steps, we observe the infection time $\tau^c$ of nodes as they subsequently become infected; for nodes $u$ that are never infected, we set the infection time $\tau_u^c=\infty$. Furthermore, for the activated nodes, we do not observe which node activated them. Therefore, the connectivity structure of the network is hidden. There exist a number of techniques that can estimate the parameters $p_{vw}$ for each edge using this observation model (Myers and Leskovec, 2010; Gomez-Rodriguez *et al.*, 2012; Netrapalli and Sanghavi, 2012; Wang *et al.*, 2012).

## 3 Collective Diffusion—CSI Model

The typical observation model used in the social networking setting assumes that one can track the status of each individual in the network and exploit this information during inference. However, this assumption rarely holds in several computational sustainability domains such as ecology, social sciences and transportation, where data identifying a single individual is not often available. We motivate this observation through the following examples.

In wildlife population modeling, the goal is to describe the occupancy pattern of habitat patches for a certain species in a fragmented landscape over a period of time (Hanski, 1999; Sheldon *et al.*, 2010). Each habitat patch $i$ can be thought of as a node in a geospatial network. A habitat patch $i$ can provide support for at most $N_i$ members of a species. This model works

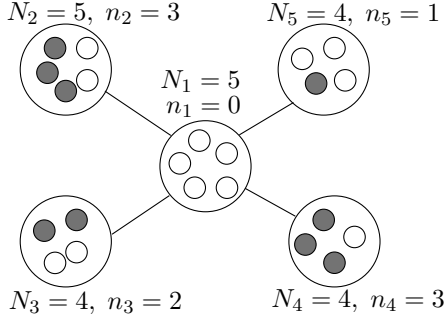

Figure 2: An example of collective diffusion in a 5-node network. Each small circle represents an individual within the larger node. The total population of a node is given as $N_i$; number of active individuals within a node shown in grey circles is denoted as $n_i$

well for many animals, but is particularly appropriate for territorial species, in which an individual or a family group defends a distinct territory within the patch for breeding and foraging. A concrete example is a species of cavity-nesting bird such as the Eastern Bluebird, who do not excavate their own cavities, but rely on those made by other species. In this case, the number $N_i$ corresponds to the number of available nest cavities. As it is difficult to track individual birds, the observed data often consists of only the number of species present in a habitat patch $i$, say $n_i$. Based on this *aggregate* observed data, we need to infer the *colonization* probability $p_{vw}$ that represents the probability that an individual from patch $v$ will colonize an unoccupied cavity in patch $w$.

Similar collective diffusion settings arise while modeling the spread of infectious diseases (Abbey, 1952; Halloran *et al.*, 2010). The region under observation is divided into multiple sub-communities. For example, the CDC in the United States reports flu data for 10 federal health regions as shown in Fig. 1. The observed data consists of the total number of infected individuals in a sub-community and the time data is collected (the particular week of the year). Therefore, the need to model and reason with *aggregate* data motivates the development of the following *Collective-Susceptible-Infected* (CSI) model.

### 3.1 The CSI Model

In the CSI model, we are given a graph $G = (V, E)$. Each node $i$ in the graph represents a sub-community of individuals. A node $i$ can support a maximum of $N_i$ individuals. Each individual can be either *active* or *inactive*. A complete observed cascade $c$ is the collection of nodes and the infection time $\tau_i$ and the number of individuals $n_i$ that are activated for each node $i$. For simplicity, we refer to $n_i$ as node $i$'s *activation level*. We call a node $i$ active if $n_i \geq 1$. That is, it has at least one active individual. A CSI cascade proceeds in a similar manner as described in Sec. 2:

- We start with an initial set $S_0$ of active nodes called *seeds* in the network. Each active node has an activation level $1 \leq n_i \leq N_i$. The process then unfolds in discrete time steps.

- When a node $j$ first becomes active, it is given a single chance to activate each currently inactive neighbor $i$. Each *active* individual in node $j$ can activate an *inactive* individual in node $i$ with probability $p_{ji}$. Whether node $j$ succeeds or fails in activating any individual in node $i$, it cannot make further attempts to activate $i$ in the future.

- If a currently inactive node $i$ has multiple newly active neighbors, their activation attempts of $i$ are sequenced arbitrarily.

- Node $i$, once active with activation level $n_i$, remains active with the same activation level for the entire diffusion process.

As also highlighted in Sec. 2, we can model *non-progressive* cascades in which the activation level of nodes change, such as the changing population of species in a habitat patch with time, by using the above diffusion process in a time-indexed layered graph (Kempe *et al.*, 2003).

A critical issue to address in such a collective diffusion process is to address how many individuals become active in a currently inactive node $i$. Let us assume that the current time step is $t$. Consider a single individual $i_m$ within the node $i$. Let $X(t)$ denote the set of all newly activated nodes at time $t$: $X(t) = \{i \in V : \tau_i = t\}$. The probability that $i_m$ is not active given the status of its neighbors is given as:

$$P(i_m = 0 \text{ at time } t \mid X(t-1)) = \prod_{j \in X(t-1)} (1 - p_{ji})^{n_j}$$

Therefore, the probability that individual $i_m$ is active is given as:

$$P(i_m = 1 \text{ at time } t \mid X(t-1)) = 1 - \prod_{j \in X(t-1)} (1 - p_{ji})^{n_j}$$

As the individuals within the node $i$ are identical, one can think of the process of determining the number of active individuals $n_i$ within the node $i$ as sampling from the following Binomial distribution:

$$P(n_i = \text{active at time } t \mid X(t-1)) = \frac{N_i!}{n_i!(N_i - n_i)!}$$
$$\left(1 - \prod_{j \in X(t-1)} (1 - p_{ji})^{n_j}\right)^{n_i} \prod_{j \in X(t-1)} (1 - p_{ji})^{n_j(N_i - n_i)}$$
(1)

Based on this understanding of the CSI model, we are now ready to describe the maximum likelihood formulation of the parameter learning problem.

### 3.2 Parameter Learning for CSI Model

Let $\mathcal{C}$ denote the set of all observed cascades. For each cascade $c \in \mathcal{C}$, we only observe which nodes are active at each time step and their activation level. We do not observe how a node got infected, which particular individuals within a node are active/inactive or the underlying connectivity structure of the network. The goal is to learn the parameters $p_{ij}$ for each pair of nodes $i, j \in V$. Our approach is based on maximizing the likelihood of the observed data. Similar maximum likelihood (ML) based approaches have been used in (Myers and Leskovec, 2010; Netrapalli and Sanghavi, 2012). However, previous approaches are not applicable to the collective variant which we address.

Let $A$ denote the matrix of activation probabilities. Let $c$ denote a particular cascade $c \in \mathcal{C}$. Let $\mathcal{I}^c$ denote the set of nodes that become activated at some point in cascade $c$; $\mathcal{U}^c$ denote the nodes that remain unactivated. The probability of the observed cascades is:

$$P(\mathcal{C};A) = \prod_{c \in \mathcal{C}} \left[ \left( \prod_{i \in \mathcal{I}^c} P\big(n_i^c \text{ active at time } \tau_i^c | X^c(\tau_i^c - 1) \big) \right. \right.$$
$$P\big(\text{Node } i \text{ inactive before time } \tau_i^c | X^c(t < \tau_i^c - 1) \big) \bigg)$$
$$\left. \left( \prod_{i \in \mathcal{U}^c} P(\text{Node } i \text{ always inactive}) \right) \right] \quad (2)$$

where $X^c(t < \tau_i - 1)$ denotes the set of all nodes that were activated before time $\tau_i^c - 1$ in a cascade $c$. The first term in the above expression is given in Eq. (1). We write the expressions for the remaining two terms for a particular cascade $c$ as follows:

$$P\big(\text{Node } i \text{ inactive before time } \tau_i^c | X^c(t < \tau_i^c - 1)\big) = \prod_{j \in V : \tau_j^c < \tau_i^c - 1} (1 - p_{ji})^{n_j^c N_i} \quad (3)$$

The probability that node $i$ never became activated is similarly given as:

$$P\big(\text{Node } i \text{ always inactive}\big) = \prod_{j \in V : \tau_j^c < \infty} (1 - p_{ji})^{n_j^c N_i} \quad (4)$$

The maximum likelihood problem entails finding the activation probability matrix $A$ that maximizes the following:

$$\max_A \log P(\mathcal{C}; A) \quad (5)$$

We can easily see that maximizing the above log-likelihood can be performed independently for each node $i$ in the network, which makes the approach highly scalable. Therefore, the optimization problem for a particular node $i$ is given as:

$$\max_{\{p_{ji}\}} \sum_{c \in \mathcal{C}: \tau_i^c < \infty} n_i^c \log\left(1 - \prod_{j \in X^c(\tau_i^c - 1)} (1 - p_{ji})^{n_j^c}\right) +$$
$$\sum_{c \in \mathcal{C}: \tau_i^c < \infty} \sum_{j \in X^c(\tau_i^c - 1)} n_j^c (N_i - n_i^c) \log(1 - p_{ji}) +$$
$$\sum_{c \in \mathcal{C}: \tau_i^c < \infty} \sum_{j \in V : \tau_j^c < \tau_i^c - 1} n_j^c N_i \log(1 - p_{ji}) +$$
$$\sum_{c \in \mathcal{C}: \tau_i^c = \infty} \sum_{j \in V : \tau_j^c < \infty} n_j^c N_i \log(1 - p_{ji}) \quad (6)$$

The above optimization is not convex and thus, direct optimization may not produce optimal solutions. We next show how to make the above problem convex by using a change of variables similar in spirit to the approach in (Myers and Leskovec, 2010). Let us introduce the following substitutions:

$$q_{ji} = 1 - p_{ji} \quad (7)$$
$$\gamma_i^c = 1 - \prod_{j \in X^c(\tau_i^c - 1)} q_{ji}^{n_j^c} \quad (8)$$
$$\hat{q}_{ji} = \log q_{ji} \quad (9)$$
$$\hat{\gamma}_i^c = \log \gamma_i^c \quad (10)$$

The new equivalent optimization problem is given as:

$$\max_{\{\hat{q}_{ji}, \hat{\gamma}_i^c\}} \sum_{c \in \mathcal{C}: \tau_i^c < \infty} \left\{ n_i^c \hat{\gamma}_i^c + \sum_{j \in X^c(\tau_i^c - 1)} n_j^c (N_i - n_i^c) \hat{q}_{ji} \right.$$
$$\left. + \sum_{j \in V : \tau_j^c < \tau_i^c - 1} n_j^c N_i \hat{q}_{ji} \right\} + \sum_{c \in \mathcal{C}: \tau_i^c = \infty} \sum_{j \in V : \tau_j^c < \infty} n_j^c N_i \hat{q}_{ji} \quad (11)$$

$$\text{s.t. } \exp\left(\hat{\gamma}_i^c\right) + \exp\left(\sum_{j \in X^c(\tau_i^c - 1)} n_j^c \hat{q}_{ji}^c\right) \leq 1 \; \forall c : \tau_i^c < \infty \quad (12)$$

$$\hat{q}_{ji} \leq 0 \; \forall j \in V \quad (13)$$
$$\hat{\gamma}_i^c \leq 0 \; \forall c : \tau_i^c < \infty \quad (14)$$

In the above problem, the objective function is linear in all the variables. We have represented the equality constraint in Eq. (8) using the inequality constraint (12). This is justified as the objective function is an increasing function of both $\hat{\gamma}_i^c$ and $\hat{q}_{ji}$. Therefore, at the optimal solution, there will be no slack for this constraint and Eq. (8) will hold. To make constraint (12) convex, we take log of both the sides and get an equivalent convex constraint as:

$$\log\left(\exp\left(\hat{\gamma}_i^c\right) + \exp\left(\sum_{j \in X^c(\tau_i^c - 1)} n_j^c \hat{q}_{ji}^c\right)\right) \leq 0 \quad (15)$$

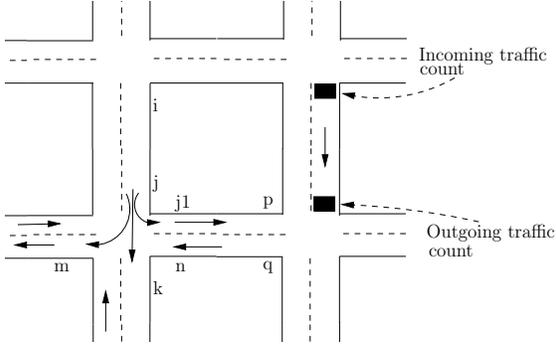

Figure 3: A road network showing data collection methodology using loop detectors that can count incoming and outgoing vehicles for a road segment

Now the optimization problem (11) is a convex problem subject to constraints (13,14,15) and therefore, can be solved optimally using off-the-shelf nonlinear solvers such as SNOPT.

As also noted in previous work (Myers and Leskovec, 2010), networks are generally sparse. To encourage sparsity, we add the following sparsity inducing penalty term to the objective function for a node $i$, with parameter $\rho > 0$:

$$-\rho \sum_{j \in V} e^{-\hat{q}_{ji}} \qquad (16)$$

The above penalty term accurately predicts edges which are not part of the underlying network. That is, $p_{ji}=0$ for such edges. However, a side effect of the penalty term is that it skews the true parameters $p_{ji}$ for other edges. Therefore, once the underlying structure of the network is discovered using this penalty term, we solve the optimization problem again to accurately recover the true parameters $p_{ji}$.

## 4 Collective Flow Diffusion Model

In the collective diffusion model of the previous section, the total number of activated individuals and nodes increases as the underlying contagion spreads through the network. In applications such as traffic flow modeling, given certain input traffic through network entry points, one is interested in modeling how the traffic flow diffuses through the road network. Thus, *flow conservation* is observed for each node of the network. To address such scenarios, we present the *collective flow diffusion*s (CFD) model. The CFD model can also be interpreted as a Markov chain, and as highlighted in later sections, is a special case of collective graphical models (CGMs) (Sheldon and Dietterich, 2011). This connection allows for adapting inference strategies for CGMs to the CFD model.

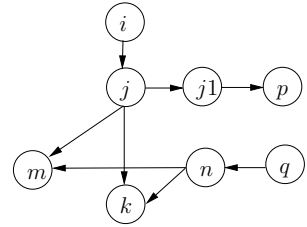

Figure 4: Equivalent network representation for the road network shown in Figure 3, includes only marked locations.

Consider a road network as shown in Fig. 3. A key learning problem in such traffic networks is estimating the *turn probabilities* for each road segment of this network. That is, given a road segment $(i, j)$ as shown in Fig. 3, we want to estimate what fraction of outgoing traffic from location $j$ goes straight, turns right and turns left over a period of time $T$. Turn probabilities are essential to model the traffic flow in several traffic simulators (Nguyen *et al.*, 1997; Thiebaux *et al.*, 1999) and are a crucial measure that determine the macroscopic properties of the traffic flow such as the congestion level, origin-destination matrix, among others. Several popular analytical models of traffic flow such as the cell transmission model (Daganzo, 1994) are based on the assumption that turn probabilities are known a priori for each location.

In several urban traffic networks, aggregate data in the form of vehicle count is already collected for each road segment using inductive-loop traffic detectors. The main data requirement in our work is the availability of aggregate *incoming* and *outgoing* traffic for a road segment for the total time duration $T$. For example, black rectangles in Figure 3 show the places where we require aggregate vehicle count. This assumption can be relaxed in principle by treating unavailable vehicle counts as missing data. For now, we handle the simpler case where such traffic counts are available for each road segment. Note that determining turn probabilities from this data is not trivial as we do not observe how much traffic is forwarded to each adjacent link.

### 4.1 Network and Data Representation

We present the CFD model in the context of traffic networks, but this model applies to any setting where flow is conserved. Each location in the road network is a node in our graph representation. For example, we create one node for each location $j$, $k$ and $m$, among others, for the road network in Figure 3. *Directed edges* model road links along with traffic direction. For example, there are directed edges $(j, j1)$, $(j, k)$ and $(j, m)$. The node $j1$ has a single outgoing link to node $p$ for the example in Figure 3. We call location nodes which receive incoming traffic, such as nodes $i$, $j1$ and $m$ as *inflow* nodes. The nodes where outgoing traffic count is recorded, such as node $j$ and $p$, are called *outflow* nodes. Figure 4 shows a part of the network representation for the road network in Figure 3.

**Observed Data:** We observe, for each *inflow* node $i$ in the network, the total incoming traffic count $n_T(i)$ after $T$ time steps. For each *outflow* node $o$, we observe the total outgoing traffic count $n_{T-1}(o)$ after $T-1$ time steps.

### 4.2 Complete Data Likelihood

We have the following flow conservation relations for different nodes in the network:

$$n_T(i) = \sum_{o \in \text{Nb}(i)} n_{T-1}(o, i) \qquad (17)$$

where $\text{Nb}(i)$ denotes adjacent *outflow neighbors* $o$ of the inflow node $i$ that send a total of $n_{T-1}(o, i)$ vehicles to node $i$ after $T-1$ time steps. In this section, we are assuming that $n_{T-1}(o, i)$ is also observed; the results for this simpler case will pave the way an outline of the Expectation-Maximization algorithm in the next section for the case when only total incoming and outgoing counts are observed. Let $\mathcal{O}$ denote the set of all outflow nodes in the network and $\mathcal{I}$ denote the set of inflow nodes. For each outflow node $o$, the flow conservation is given as:

$$n_{T-1}(o) = \sum_{i \in \text{Nb}'(o)} n_{T-1}(o, i) \qquad (18)$$

where $\text{Nb}'(o)$ denotes adjacent *inflow neighbors* $i$ of the outflow node $o$ that can receive traffic from $o$. The turn probabilities $p_{oi}$ are defined for each pair $(o, i)$ of adjacent outflow and inflow nodes. They intuitively represent the probability that a vehicle at the node $o$ will turn to node $i$. The complete data joint probability is given as:

$$P(\boldsymbol{n}; \{p_{oi}\}) = \prod_{o \in \mathcal{O}} \left[ \frac{n_{T-1}(o)!}{\prod_{i \in \text{Nb}'(o)} n_{T-1}(o, i)!} \prod_{i \in \text{Nb}'(o)} p_{oi}^{n_{T-1}(o, i)} \right]$$
(19)

Subject to the following constraints:

$$\sum_{i \in \text{Nb}'(o)} p_{oi} = 1 \qquad \forall o \in \mathcal{O} \qquad (20)$$

$$\sum_{i \in \text{Nb}'(o)} n_{T-1}(o, i) = n_{T-1}(o) \quad \forall o \in \mathcal{O} \qquad (21)$$

$$\sum_{o \in \text{Nb}(i)} n_{T-1}(o, i) = n_T(i) \qquad \forall i \in \mathcal{I} \qquad (22)$$

The meaning of the constraints is that the probability is zero when the observed data do not satisfy flow conservation. Intuitively, the expression in Eq. (19) is a product of multinomial distributions, one for each outflow node $o$, where one can imagine performing $n_{T-1}(o)$ trials which lead to success in exactly one of the categories in the set $\text{Nb}'(o)$. This joint-probability describes a single cascade of flow diffusion and can be easily generalized to multiple independent cascades by using the i.i.d. assumption. Given the complete observed data $\boldsymbol{n}$, estimating the turn probabilities involves solving the following optimization problem subject to constraint (20), which is equivalent to estimating the parameters of a multinomial distribution.

$$\max_{\{p_{oi}\}} \sum_c \log P(\boldsymbol{n}; \{p_{oi}\}) \qquad (23)$$

where $c$ denotes a single complete cascade.

### 4.3 Inference With Missing Data

According to the observation model in Section 4.1, the variables $n_{T-1}(o, i)$ are not observed for any location pairs. Therefore, the approach of Section 4.2 cannot be applied directly. However, recently inference approaches to address missing data in collective graphical model settings have been proposed (Sheldon and Dietterich, 2011). Sheldon and Dietterich (2011) address the problem of generating samples from the posterior distribution of a collective graphical model by deriving an efficient Gibbs sampling algorithms that can work with hard constraints as in (21) and (22). Therefore, such a sampling strategy can be used in conjunction with the EM algorithm (Dempster *et al.*, 1977) to generate complete data samples conditioned on aggregate data for the traffic network. The M-step of the EM algorithm involves a similar optimization as in Eq. (23).

## 5 Experiments

In this section, we present the results of our inference approach for a number of synthetic and real-world data sets. We focus on the collective diffusion model of Section 3. Our diffusion simulator was implemented in JAVA and used the nonlinear programming solver SNOPT (Gill *et al.*, 2002) with AMPL interface (Fourer *et al.*, 2002) to solve the optimization problem (11). The experiments were run on a Mac Pro with a single 2.4GHz processor and 4GB RAM allocated to the solver.

For synthetic benchmarks, we generated 100, 250 and 500 node scale-free networks with the preferential attachment model similar to (Myers and Leskovec, 2010). The largest 500 node network had about 900 edges. The edges were considered bi-directed, implying 1800 edges for the 500 node network. The edge log-activation probabilities, $\ln p_{ij}$, were sampled uniformly randomly in the range $[-8, -4.6]$ for each run. A main objective of the experiments is to test the efficacy of the optimization approach of Section 3.2 w.r.t. a varying number of cascades. Ideally, one would like to learn the network structure and the edge parameters accurately with as few cascades as possible. Encouragingly, our approach is quite successful in achieving this objective as highlighted next.

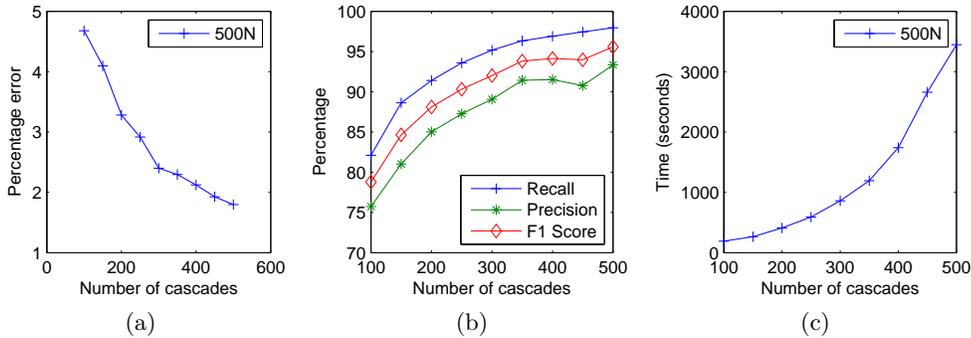

Figure 5: Error, recall, precision and timing results for the 500 node network

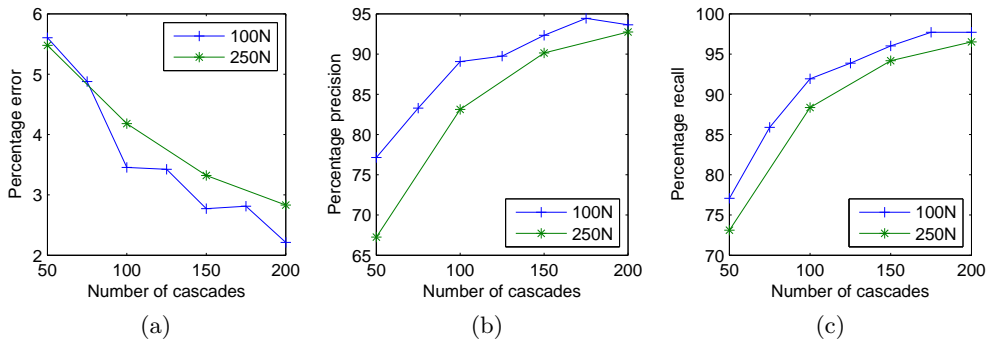

Figure 6: Error, precision and recall results for 100 and 250 node networks

Figure 5 shows the results for the largest 500 node network. Each point in the plots is the average of 5 runs. We fixed the maximum population $N_i$ of each node to 1000. To make the inference challenging, 5% of total nodes were initialized as seeds, resulting in 25 seeds for the 500 node network. The activation level of seeds was sampled uniformly from the range $[5, 25]$. The x-axis of each plot represents the number of cascades and the y-axis shows the measured property.

Figure 5(a) shows the percentage error in estimating the edge activation parameter for the set of correctly predicted edges, say $S$, calculated as:
$$error = 100 * \frac{\sum_{ij \in S} |p_{ij}^{estimate} - p_{ij}^{true}|}{\sum_{ij \in S} p_{ij}^{true}} \qquad (24)$$

In this 500 node network, even with 100 cascades, the error is quite small, around 5%. We contrast this with the setting in (Myers and Leskovec, 2010), where roughly the same number of cascades were generated as the number of nodes. Our results show that under the collective diffusion model, one can obtain good results with significantly fewer cascades. As expected, the error decreases as the number of observed cascades increases. For 500 cascades, it is around 2%, resulting in very high accuracy.

Figure 5(b) shows the precision, recall and the $F_1$ score with varying number of cascades. For 100 cascades, the precision and recall are 75% and 82% respectively. It is encouraging that we can get reasonable results even with very few cascades. Furthermore, both the precision and recall increase sharply w.r.t. the number of cascades. The $F_1$ score is around 90% for 250 cascades. This shows that our approach is particularly effective in utilizing additional data. For 500 cascades, the $F_1$ score is already 95% resulting in very high accuracy and precision in estimating the original network.

Figure 5(c) shows the total runtime of our approach which includes the time to generate cascades and solve the optimization problem. The runtime as expected increases w.r.t. the number of cascades. It takes about an hour to solve the largest 500 cascade instance. A main contributing factor to the runtime is the sharp increase in the size of the convex program for each node with the number of cascades. The SNOPT solver takes about 6 minutes to solve the optimization problem for each node, the size of which is about 3MB in the AMPL format. However, the good news is that once the cascades are generated, one can drastically reduce the runtime by solving the optimization problem for each node independently. Therefore, there is a dramatic potential for speedup by using cloud computing or multicore machines.

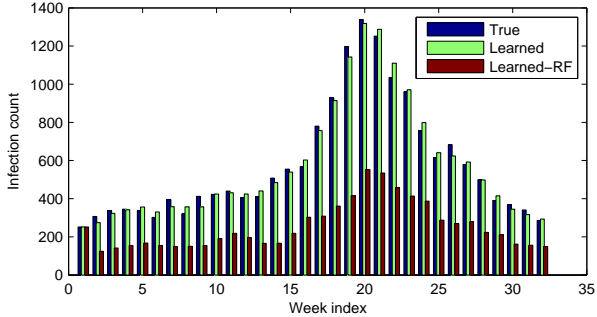

Figure 7: Comparison of true infection count, predicted counts using our approach ('Learned') and predicted counts using the Reed-Frost model ('Learned–RF') for the US Federal region 1

Figure 6 shows the error, precision and recall results for smaller 100 and 250 node networks. The results are similar to those obtained on the 500-node networks. Our approach is able to achieve more than 90% recall and precision for both these cases, further providing a proof its efficacy.

**CDC Data:** We also tested the CSI model of Section 3 to model the spread of flu in the US for the season 2010-11. The data is made publicly available by the Centers for Disease Control and Prevention (CDC, http://www.cdc.gov). The data is available for each of the 10 Federal regions of the US as shown in Figure 1. For each region, the relevant data consists of tuples ⟨Week number, # of flu patients⟩. We consider the peak of the flu season from week 40 (October) till week 20 (May) of the next year. In this setting, we consider the graph to be fully connected with each of the 10 regions potentially able to influence every other region. As the number of flu patients varies with time, we model it using non-progressive cascades using a time indexed graph, with each layer corresponding to the particular week number. The parameter $N_i$ is the total population of all the states in region $i$.

The strength of influence of a region $i$ on region $j$ is denoted as $p_{ij}$. Intuitively, it denotes how region $i$'s flu population influences the flu spread in region $j$. As no true model describing flu spread is available, we make certain assumptions that attempt to avoid overfitting of the data and represent some intuitions about the disease spread. They are as follows. First, all the variables $p_{ij}$ are independent of the particular time of the year. That is, *inter-region* spread has same parameters for every week. This represents the intuition that travel trends that affect the inter-region spread roughly remain the same throughout the year.

Second, the flu spread probability within a region $i$ (or the *intra-region* spread) for a particular week $t$ is modeled as $p_{ii}^t$ to describe how the flu spread strength varies with the time of the year. This is justified as the flu spread depends on the intensity of the cold weather, which varies with time. Finally, instead of the intra-region spread being independent for each Federal region, we constrain them to be within a certain percentage of a base probability, that itself is an optimization variable. That is:

$$0.8 \leq \frac{p_{ii}^t}{p_{\text{ref}}^t} \leq 1.2 \ \forall i, \ \forall t \qquad (25)$$

where $p_{\text{ref}}^t$ is the base flu spread probability for a particular time $t$ and is itself an optimization variable. The main effect of this constraint is that it couples the intra-region spread probability of all the regions. This constraint further attempts to avoid the overfitting of data. Finally, the main variables to estimate are the *inter-region* spread probabilities $p_{ij}$ for each pair of 10 regions, the *intra-region* spread probabilities $p_{ii}^t$ for each week $t$ and region $i$, and the base spread probability $p_{\text{ref}}^t$ for each week $t$.

We also note that a similar model to ours is used to model disease spread (Halloran *et al.*, 2010; Abbey, 1952). In particular, the Reed-Frost (RF) model (Halloran *et al.*, 2010; Abbey, 1952) is very similar to the CSI model. The key advantage of the CSI model is that it can model and learn the influence of nodes on each other. The RF model does not allow such inter-region effects and thus, its parameters are much simpler to estimate.

We first compare the accuracy of our model and the RF model. For the RF model, we include the constraint (25), otherwise it is trivial to fit the data with almost 100% accuracy by adjusting the intra-region to track observed flu intensity, which represents overfitting of data. The results for our model are quite encouraging. The average error in predictions using our model is only 3.8% for all the regions and weeks. The minimum error is 1.15% for region 5. The maximum error is 10% for region 7. For the RF model, the average error is 31%. This confirms our modeling assumption that inter-region spread is crucial to take

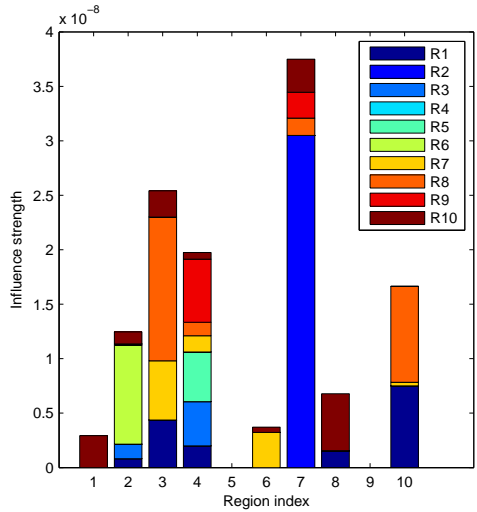

Figure 8: Inter-region influence visualization (best viewed in color). The x-axis denotes a particular region, y-axis denotes the strength of its influence on other regions using the color coding scheme on the right

into account. The accuracy of the RF model, which does not model the inter-region spread, suffers significantly. Figure 7 shows the weekly predictions for our model, the RF model and the true observed data for region 1. We can easily see the difference in the accuracy of our model and the RF model.

Figure 8 shows a visualization of the inter-region influence $p_{ij}$ for all the 10 regions. We note that as the true model for flu spread is not known, we must be careful in interpreting this result. Regardless, this plot depicts several trends that are intuitively correct and provide a justification to some of our modeling assumptions. Regions 2, 3 and 4 are the ones most responsible for spreading the flu to other regions, indicated by the number of stacks for each of them. This is intuitively justified as these regions represent the North-East and the South-East of the US (see Figure 1) and are known to have strong flu season due to intense cold weather. Region 9, which includes California, is generally known to have a weak flu season and this is reflected in Figure 8, where region 9 has zero influence on other regions.

We also note that our post-hoc analysis can be useful for health providers to better prepare for the flu season next year. The model we presented can precisely estimate the strength of flu spread for different regions and at different time of the year. This knowledge can help health care providers to prepare for contingencies for the future flu season. Currently, we only modeled the observed data for a single flu season. Predicting future flu seasons on a weekly or bi-weekly basis based on past season's data remains an important future work and will require additional analysis and inputs.

## 6 Conclusion

In several computational sustainability applications including the spread of wildlife, infectious diseases and traffic flow, the observed data often consists of only collective information, without any identifiable details about individuals in the population. In our work, we presented models that generalized the standard diffusion models such as the independent cascade model to collective settings. We motivated such collective models based on applications in ecology, infectious disease spread and transportation. We also developed scalable convex optimization based techniques that can accurately learn the parameters, including the hidden structure of the underlying network, by observing only timestamped aggregate data. Experiments on a number of synthetic and real-world benchmarks show that our approach is highly accurate and can recover the hidden structure for large networks with high precision and recall even with limited observed data.

Our future work includes further exploration of inference based techniques for modeling the traffic flow and disease spread. Addressing both these applications can create a significant practical impact. Further investigation of the connections we established between these domains and graphical models and optimization would certainty lead to deeper insights in modeling and decision making for these domains.

## Acknowledgements

Daniel Sheldon is supported by the National Science Foundation under Grant No. IIS-1117954.